\documentclass{crmp-l}

\usepackage{amssymb}
\usepackage{graphicx}

\numberwithin{equation}{section}

\newcommand{\tensor}[1]{\ensuremath{\stackrel{=}{#1}}}

\begin{document}

\title[Molecular Alignment and Orientation]{Molecular Alignment and
  Orientation: \\ From Laser-Induced Mechanisms to Optimal Control}

\author{Osman Atabek}
\address{Laboratoire de Photophysique Mol\'{e}culaire du CNRS \\ B\^{a}timent
  213 \\ Universit\'{e} Paris-Sud \\ 91405 Orsay, France}
\email{osman.atabek@ppm.u-psud.fr}

\author{Claude M. Dion}
\address{CERMICS \\ \'{E}cole Nationale des Ponts et Chauss\'{e}es \\ Cit\'{e}
  Descartes, Champs-sur-Marne \\ 77455 Marne-la-Vall\'{e}e, France}
\email{claude.dion@physics.org}

\thanks{This work benefited from the financial support of an
  \emph{Action Concert\'{e}e Incitative Jeunes Chercheurs} of the French
  Ministry of Research and from a grant of computer time from the
  \emph{Institut du D\'{e}veloppement et des Ressources Informatiques
    Scientifiques} of the CNRS}

\subjclass[2000]{}

\begin{abstract}
  Genetic algorithms, as implemented in optimal control strategies,
  are currently successfully exploited in a wide range of problems in
  molecular physics. In this context, laser control of molecular
  alignment and orientation remains a very promising issue with
  challenging applications extending from chemical reactivity to
  nanoscale design. We emphasize the complementarity between basic
  quantum mechanisms monitoring alignment/orientation processes and
  optimal control scenarios. More explicitly, if on one hand we can
  help the optimal control scheme to take advantage of such mechanisms
  by appropriately building the targets and delineating the parameter
  sampling space, on the other hand we expect to learn, from optimal
  control results, some robust and physically sound dynamical
  mechanisms.  One of the basic mechanisms for alignment (i.e.,
  molecular axis parallel to field polarization) is related to the
  pendular states accommodated by the molecule-plus-field effective
  potential. The laser control of alignment can be reached by an
  adiabatic transport of an initial isotropic rotational state on some
  pendular state trapping the molecule in well-aligned geometries.
  Symmetry breaking mechanisms are to be looked for when orientation
  (i.e., molecular axis in the same direction as field polarization)
  is the goal of laser control.  Two mechanisms are considered.  The
  first is based upon an asymmetric pulse combining a frequency
  $\omega$ and its second harmonic $2\omega$ resonant with a
  vibrational transition.  A much more efficient mechanism is the
  so-called ``kick'' that a highly asymmetric sudden pulse can impart
  to the molecule.  Half-cycle pulses, within the reach of current
  experimental technology, are among good candidates for producing
  such kicks.  Very interestingly, an optimal control scheme for
  orientation, based on genetic algorithms, also leads to a sudden
  pulsed field bearing the characteristic features of the kick
  mechanism.  Optimal pulse shaping for very efficient and
  long-lasting orientation, together with robustness with respect to
  temperature effects, are among our future prospects.
\end{abstract}

\maketitle

\section{Introduction}

Laser-induced molecular alignment and orientation are challenging
control issues with a wide range of applications, extending from
chemical reactivity to nanoscale
design~\cite{align:charron94b,manip:dey00,align:poulsen02,focus:seideman97b,%
  focus:stapelfeldt97,orient:tenner91}.  They address molecular
manipulation, involving external angular degrees of freedom, aiming at
a parallel positioning of the molecular axis with respect to the laser
polarization vector (\emph{alignment}) or, even more demanding, with a
given direction (\emph{orientation}).  One of the basic motivations
remains the drastic increase of reactive cross sections, highly
sensitive to stereodynamical effects (frontal collisions).
Laser-induced isomerization~\cite{iso:dion96,iso:kurkal01}, isotope
separation~\cite{align:charron94b}, molecular
trapping~\cite{focus:seideman97a}, or surface processing and
catalysis~\cite{manip:dey00,focus:seideman97b,orient:tenner91} are
some of the illustrations of the widely growing interest in molecular
manipulation.  

The most common idea is to excite a broad rotational band $\Delta J$
in order to recover a narrow angular distribution $\Delta \theta$,
through the celebrated relation $\Delta J \Delta \theta \sim \hbar$
($J$ being the total angular momentum and $\theta$ the molecule-laser
angle).  Our approach to this problem is, however, different.  The
emphasis is put on the depiction and study of some laser-induced
dynamical mechanisms that are relevant to the alignment/orientation
process: pendular states in high-frequency fields; two-photon
(fundamental and second-harmonic) excitations; the kick mechanism.
The question of how to take advantage of such mechanisms in a control
scheme is brought up only in a second step.  Adiabatic transport from
an isotropic distribution to a pendular state; two IR laser pulses
with frequencies in a ratio 2, the second harmonic being in resonance
with a vibrational transition; very asymmetric and short-duration
(sudden) half-cycle pulses imparting sudden kicks to the molecule, are
some of the scenarios which are depicted and discussed in the
following.  Finally, in a third step, this knowledge is fully
exploited in optimal control schemes, not merely for initial guesses
in parameter sampling, but mainly to get a better understanding and
interpretation of otherwise black-box type calculations.

The paper is organized as follows.  Section~\ref{sec:align} is devoted
to intense laser alignment of the HCN molecule, through pendular
states as the elementary mechanism and adiabatic/sudden transport as
the control strategy.  Section~\ref{sec:w2w} presents the two-color
orientation of HCN by a parity mixing scenario in the rotational
distribution.  The kick mechanism with a half-cycle pulse sudden
excitation scenario is examined in section~\ref{sec:hcp}, as
illustrated on the LiCl molecule.  Sudden-excitation-based optimal
control scenarios, with a thorough analysis of the role of different
orientation criteria, are analyzed, using HCN and LiF, in
section~\ref{sec:optimal}.

\section{Laser-controlled alignment} \label{sec:align}

\subsection{The model}

\begin{figure}
\includegraphics{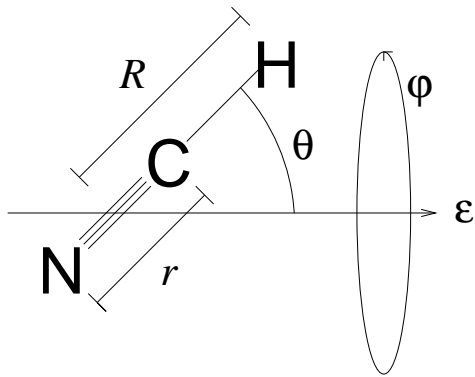}
\caption{The linear HCN molecule positioned, with respect to the
  polarization axis $\mathcal{E}$ of the laser, by its polar $\theta$
  and azimuthal $\varphi$ angles.  Internal stretching coordinates are
  collectively noted $\mathbf{R} = (R,r)$.}
\label{fig:model}
\end{figure}

The linear HCN molecule, taken as an illustrative example, is
described by its internal stretching motions (collectively labelled
$\mathbf{R}$) and external rotational motions described by its polar
$\theta$ and azimuthal $\varphi$ angles with respect to the laser
polarization vector (see figure~\ref{fig:model}). 
The Hamiltonian involves a single Born-Oppenheimer potential curve
$V(\mathbf{R})$ (the molecular electronic ground state) and the
radiative coupling $\hat{H}_{\mathrm{rad}}(\mathbf{R}, \theta,
\varphi, t)$ given by the scalar product of the field-induced dipole
moment $\mbox{\boldmath $\mu$}(\mathbf{R})$ with the electric field
amplitude $\mbox{\boldmath $\mathcal{E}$}(t)$, i.e.,
\begin{equation}
\hat{H}(\mathbf{R}, \theta, \varphi, t) = \hat{T}_{\mathbf{R}} +
\frac{\hbar^2}{2 I(\mathbf{R})} \hat{J}^{2} + V(\mathbf{R}) +
\hat{H}_{\mathrm{rad}}(\mathbf{R}, \theta, \varphi, t),
\end{equation}
where $\hat{T}_{\mathbf{R}}$ is the kinetic energy operator in
$\mathbf{R}$, $\hat{J}^{2}$ the angular momentum operator, and
$I(\mathbf{R})$ the moment of inertia of the molecule.  The dipole
moment itself is taken  as the sum of a permanent dipole
$\mbox{\boldmath $\mu$}_{0}$ and a polarizability $\tensor{\alpha}$
\begin{equation}
\mbox{\boldmath $\mu$}(\mathbf{R},t) = \mbox{\boldmath $\mu$}_{0} +
\frac{1}{2} \tensor{\alpha} \cdot \mbox{\boldmath $\mathcal{E}$}(t),
\end{equation}
such that
\begin{equation}
\hat{H}_{\mathrm{rad}}(\mathbf{R}, \theta, \varphi, t) = - \mu_{0}
\mathcal{E}(t) \cos\theta - \frac{1}{2} \mathcal{E}^{2}(t) \left(
  \alpha_{\parallel} \cos^{2}\theta + \alpha_{\perp} \sin^{2}\theta
\right).
\end{equation}
In the last equation, we have made use of the fact that, for a linear
molecule, $\tensor{\alpha}\ = \mbox{Diag} \left\{ \alpha_{xx},
  \alpha_{yy}, \alpha_{zz} \right\}$, with $\alpha_{xx} = \alpha_{yy}
\equiv \alpha_{\perp}$ and $\alpha_{zz} \equiv \alpha_{\parallel}$,
when the $z$ axis is chosen to be collinear with the molecular axis.
Due to cylindrical symmetry and since the laser-molecule coupling term
does not depend explicitly on $\varphi$, the motion on $\varphi$ can
be considered separately and is no longer taken into account.

The laser pulses are of the form $\mathcal{E}(t) = \mathcal{E}_{0}(t)
\cos \left( \omega t + \phi \right)$, with $\mathcal{E}_{0}(t)$ the
time varying envelope of the electric field of frequency $\omega$ and
phase $\phi$.  We will consider laser pulses of a maximum intensity of
the order of $10^{13}\ \mathrm{W/cm}^{2}$, to avoid ionization damage
to the molecule.  The frequency $\omega$ is taken in the infra-red
region, so it is considerably higher than the rotational frequency of
the molecule.

The dynamics of the molecular system are then obtained by solving
numerically the time-dependent Schr\"{o}dinger equation
\begin{equation}
i \hbar \frac{\partial}{\partial t} \psi(\mathbf{R},\theta,t)
= \hat{H}(\mathbf{R},\theta,t)
\psi(\mathbf{R},\theta,t) 
\end{equation}
(on a discretized grid using the split-operator
method~\cite{split:adb1991,fft:feit82,split:fleck76}), starting from
an initial wave function $\psi(\mathbf{R},\theta,t=0)$ taken, in most
cases, as the ground (rotationally isotropic) state of the molecule.

\subsection{Pendular states}

In this high-frequency regime, two different motions are considered: a
fast motion associated with the electric field $\omega t$ and a slow
motion associated with the molecular rotation described by $\theta$.
To simplify the presentation, we freeze the molecular internal
motions, assuming a rigid rotor behavior.  The eigenvectors of the
slow motion $\xi_{nJ} (\theta)$ obey, after averaging over the fast
motion within an optical period (and some high-frequency
approximations) a Schr\"{o}dinger equation given
by~\cite{floquet:keller00a}
\begin{eqnarray}
  \left[ \frac{\hbar^{2}}{2 I} \hat{J}^{2} + \frac{\hbar^{2}}{2 I}
    \frac{M^{2}}{\sin^{2} \theta} + \left( \Delta\alpha +
      \frac{\mu_{0}^2}{I \omega^{2}} \right) \frac{\mathcal{E}_{0}}{4}
    \sin^{2} \theta + \frac{\Delta\alpha^{2} \mathcal{E}_{0}^{4}}{256 I
      \omega^{2}} \sin^{2} 2\theta  \right] \xi_{nJ} (\theta) \nonumber \\
  = \left( \lambda_{n} - J\hbar\omega + \frac{\alpha_{\parallel}
  \mathcal{E}_{0}^{2}}{4} \right) \xi_{nJ} (\theta),
  \label{eq:hf}
\end{eqnarray}
where $\Delta\alpha = \alpha_{\parallel} - \alpha_{\perp}$.  The
$\theta$-dependent laser-induced effective potential has a
$\sin^{2}\theta$ shape and supports the eigenvectors $\xi_{nJ}
(\theta)$ which are localized around $\theta = 0$ and $\pi$ (aligned
geometry), in a way which is more marked for increasing intensity.
Because of the resulting libration motion for a molecule that is in
one of these eigenstates, they are dubbed \emph{pendular
  states}~\cite{align:friedrich95a,pendul:zon75}.

It is interesting to note a close analogy with the classical Lagrange
equation of the laser-driven rigid rotor under the combined effect of
the permanent dipole and the polarizability~\cite{align:dion99},
\begin{equation}
  \label{eq:lagrange}
  I \ddot{\theta} \simeq \frac{M^{2}}{I} \frac{\cos \theta}{\sin^{3}
  \theta} - \left( \Delta\alpha + \frac{\mu_{0}^2}{I \omega^{2}}
  \right) \frac{\mathcal{E}_{0}^{2}}{4} \sin 2\theta -
  \frac{\Delta\alpha^{2} \mathcal{E}_{0}^{4}}{128 I \omega^{2}} \sin
  4\theta.  
\end{equation} 
The classical force of the right-hand-side of
equation~(\ref{eq:lagrange}) is nothing but the gradient of the
effective potential of the Hamiltonian version,
equation~(\ref{eq:hf}).  This classical analogy leads, upon a change
of variable $\Theta = 2\theta$, to the well-known equation of the
pendulum,
\begin{equation}
  \label{eq:aligndyna}
  \ddot{\Theta} + \Omega^{2} \sin \Theta = 0 ,
\end{equation}
that librates with small oscillations $\theta$ around the laser
polarization axis with a frequency
\begin{equation}
  \label{eq:Omega}
  \Omega = \frac{1}{\sqrt{I}} \left( \frac{1}{2} \frac{\mu_{0}^{2}
    \mathcal{E}_{0}^{2}}{I \omega^{2}} + \frac{1}{2} \Delta\alpha
    \mathcal{E}_{0}^{2} \right)^{1/2},  
\end{equation}
where the polarizability remains the leading term as soon as the laser
frequency exceeds 315~cm$^{-1}$ (for the HCN molecule), whatever the
intensity.

\subsection{Control scheme}

The question of how to take advantage of the pendular mechanism to
control molecular alignment can be answered by referring to adiabatic
vs sudden transport dynamics, starting from an initial isotropic
distribution ($J=M=0$, as prepared, for instance, by laser cooling
methods~\cite{mf:bahns00}).

\begin{figure}
\includegraphics{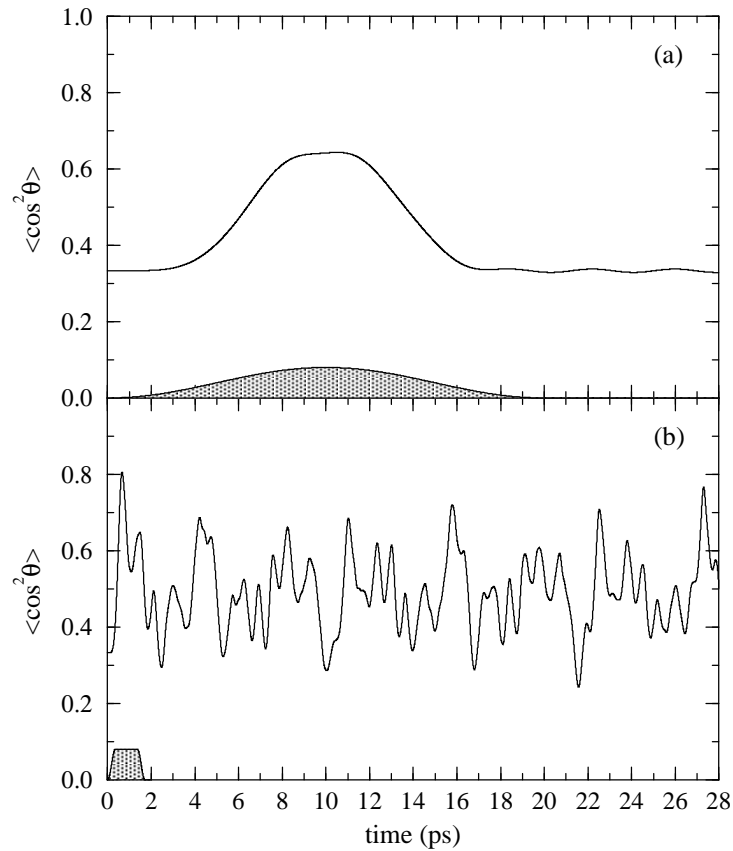}
\caption{Alignment dynamics of the HCN molecule submitted to (a) a
  long, adiabatic laser pulse of duration 10~ps and intensity
  $10^{12}\ \textrm{W/cm}^{2}$; (b) a short, sudden pulse of 1.7~ps
  duration and of intensity $10^{13}\ \textrm{W/cm}^{2}$.  The pulse
  envelopes are schematically represented by the shaded areas.}
\label{fig:adia}
\end{figure}

The adiabatic switching of a laser with rise and fall times of 10~ps
(same order as the molecular rotational period) and an intensity of
$10^{12}\ \mathrm{W/cm}^{2}$ allows an adiabatic transport on a single
(lowest) pendular state with a good alignment during the excitation.
Figure~\ref{fig:adia}(a) illustrates the alignment dynamics through the
expectation value of $\cos^2 \theta$, i.e.,
\begin{equation}
  \left\langle \cos^{2} \theta \right\rangle (t) = \int_{0}^{\pi} \cos^{2}
  \theta \left| \psi(\theta,t) \right|^{2} \sin\theta \, d\theta.
\end{equation}
The higher the value, the better the alignment, with $\left\langle
  \cos^{2} \theta \right\rangle = 1/3$ corresponding to an isotropic
distribution.  An alignment of $\left\langle \cos^{2} \theta
\right\rangle \approx 0.64$ is achieved during the pulse but, as
expected, the laser extinction adiabatically transports the pendular
state back to the isotropic ($J=M=0$) state.

To achieve post-pulse alignment, one may proceed with sudden
excitations.  Figure~\ref{fig:adia}(b) shows the effect of a sudden
pulse of 1.7~ps duration, which excites several pendular states with a
specific distribution (for a field frequency resonant with HCN
internal vibrations).  The particular $J$ distribution reached is such
that alignment remains even after the pulse is off, $\left\langle
  \cos^{2} \theta \right\rangle (t)$ showing small amplitude
oscillations around a value of $0.5$.

\section{Two-color laser control of orientation (Parity breaking
  scenarios)} \label{sec:w2w}

\begin{figure}
\includegraphics{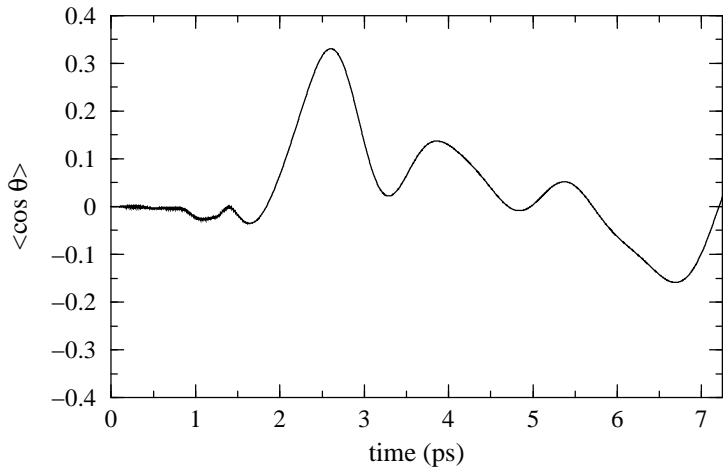}
\caption{Orientation dynamics of the HCN molecule submitted to a laser
  pulse (1.7~ps duration) combining fields of frequency $\omega$ and
  $2\omega$, the latter being in resonance with the $v = 0 \rightarrow
  1$ vibrational transition of the C--H stretch.}
\label{fig:w2w}
\end{figure}

As compared to alignment, orientation, which imposes a given
direction, is a more challenging goal that requires symmetry breaking
in the forward and backward directions.  Two such scenarios are
considered in this work.  The first concerns the use of two-color
pulses, that is a combination of a fundamental frequency $\omega$ and
its second harmonic $2\omega$ with a resulting field
\begin{equation}
\mathcal{E}(t) = \mathcal{E}_{0}(t) \left[ \cos \omega t + \gamma \cos
  \left( 2 \omega t + \phi \right) \right],
\end{equation}
where maximum asymmetry is reached for $\gamma = 0.5$ and $\phi = 0$.
But the asymmetry we are referring to is not in the time dependence of
$\mathcal{E}(t)$, which in any case is not enough to produce
orientation as, in average over an optical period of the field, the
positive and negative contributions are canceling each other.  We are
rather referring to parity breaking in the rotational quantum numbers
$J$.  As opposite to a monochromatic field that excites $J$'s of the
same parity, two-color excitation has the potentiality to mix $J$'s of
different parity.  Actually, by taking $2\omega$ in resonance with the
$v = 0 \rightarrow 1$ vibrational transition of HCN, the absorption of
a single photon $2\omega$ excites odd $J$'s through the permanent
dipole interaction $\mu_{0} \mathcal{E}$, whereas the absorption of
two photons $\omega$ excites even $J$'s through the polarizability
interaction $\alpha \mathcal{E}^{2}$ (starting from an initial state
$J=M=0$)~\cite{orient:dion99}.  It is precisely the mixture of odd and
even $J$'s that produces, by the superposition of associated spherical
harmonics, the asymmetrical angular distributions that are looked for.
The result is shown in figure~\ref{fig:w2w}(a) as the expectation
value of $\cos\theta$, which is a measure of orientation (the higher
the absolute value, the better the orientation).
We see that orientation is actually achieved at the extinction of the
laser pulse and more efficiently a short time after the pulse is off,
with $\left\langle \cos\theta \right\rangle$ reaching $0.3$.

Orientation can also be achieved by an adiabatic transport from an
initial isotropic state to an asymmetric combination of pendular
states by a two-color excitation
scheme~\cite{orient:guerin02,orient:kanai01}.  Contrary to the result
shown in figure~\ref{fig:w2w}, such an orientation can only be
maintained while the laser is on, similarly to what is observed for
alignment in figure~\ref{fig:adia}(a).

\section{Half-cycle pulses and the kick mechanism} \label{sec:hcp}

The second symmetry breaking scenario addresses the temporal shape of
the electric field.  The most asymmetrical situation that is experimentally
reachable is the use of half-cycle pulses
(HCP's)~\cite{hcp:bucksbaum00}.  Such picosecond, far-IR
electromagnetic pulses are generated when illuminating a GaAs wafer,
in the presence of a pulsed electric field applied across the surface,
by a Ti-Sapphire laser~\cite{hcp:you93}.  Such pulses have three
interesting characteristics (as illustrated on figure~\ref{fig:hcp}):
\begin{figure}
\includegraphics{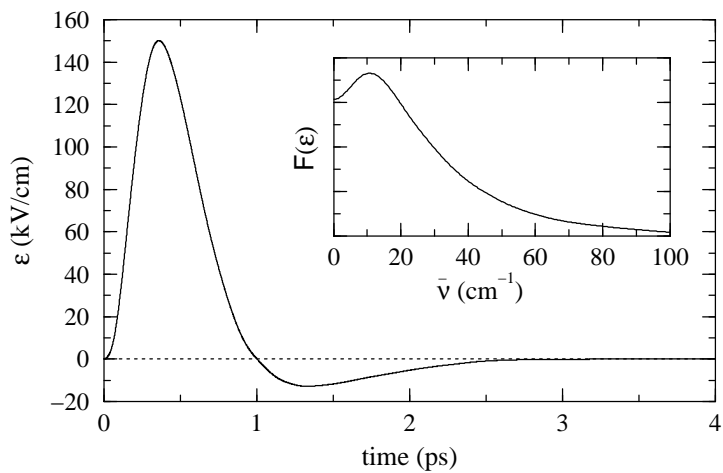}
\caption{Electric field of a half-cycle pulse as a function of time
  with its Fourier transform given in the inset.  (Adapted from
  reference~\cite{hcp:you93},  taken from C.~M. Dion, A. Keller, and
  O. Atabek, Eur. Phys. J. D \textbf{14} (2001), 249--255, copyright
  (2001) by EDP Sciences.)} 
\label{fig:hcp}
\end{figure}
\renewcommand{\theenumi}{\roman{enumi}}
\begin{enumerate}
\item their intensities remain less than $10^{8}\ \mathrm{W/cm}^{2}$,
  therefore present no risk of ionization damage on typical diatomic
  molecules; 

\item their relatively broad Fourier transform ($\sim 30\
  \mathrm{cm}^{-1}$) shows the presence of non-negligible field
  components that can induce rotational excitations;
  
\item they are very sudden with respect to rotation (their positive
  component lasts typically 1~ps) and very asymmetric (the ratio of
  the positive to the negative maximum is more than 10).
\end{enumerate}
A direct consequence is that such unipolar pulses can impart a
``kick'' to the molecule that effectively orients it in the direction
of the positive field component, the transfer of angular momentum
taking place on the timescale of the short pulse~\cite{hcp:dion01}.
The long-duration negative tail of the HCP is adiabatic enough not to
induce further noticeable changes in the dynamics.  Actually, the kick
seems the most efficient orientation mechanism up to date.  The
results are illustrated on the LiCl molecule in terms of the dynamics
of the average value of $\cos \theta$, in figure~\ref{fig:hcpkick}.
\begin{figure}
\includegraphics{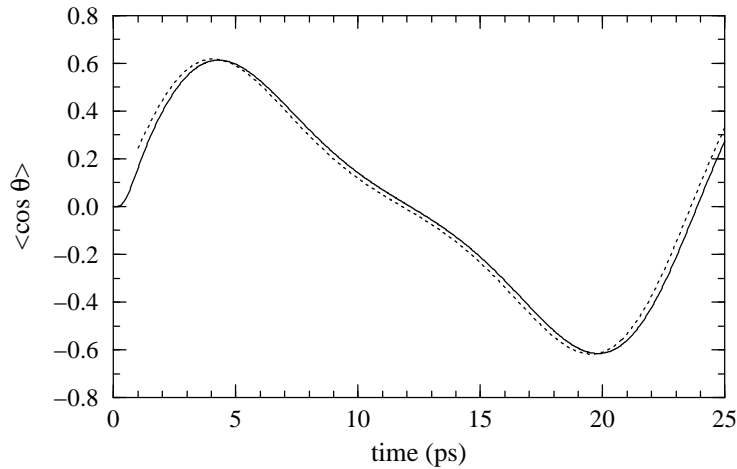}
\caption{Orientation dynamics of the LiCl molecule submitted to the
  half-cycle pulse of figure~\ref{fig:hcp} (solid line) and using a
  sudden-impact approximation (dotted line). (Taken from C.~M. Dion,
  A. Keller, and O. Atabek, Eur. Phys. J. D \textbf{14} (2001),
  249--255, copyright (2001) by EDP Sciences.)}
\label{fig:hcpkick}
\end{figure}

As in the two-color excitation scenario, the orientation remains
modest during the positive part of the pulse, but it develops later
and reaches the value $\left\langle \cos \theta \right\rangle \sim
0.5$ over a large time period of $\approx 4\ \mathrm{ps}$.  The kick
mechanism by itself can be evidenced by working out a sudden impact
model~\cite{hcp:dion01}.  Such a model is based on the short duration
($t_{\mathrm{p}} = 1\ \mathrm{ps}$) of the pulse as compared to the
molecular rotational period ($T_{\mathrm{rot}} = 24\ \mathrm{ps}$).
The calculation consists in freezing the rotational dynamics during
the time the positive part of the HCP is acting.  At the lowest order,
the approximation is valid as far as
\begin{equation}
\frac{B \hat{J}^{2} t_{\mathrm{p}}}{\hbar} \sim 0.13 J \left( J+1
\right) \ll 1,
\end{equation}
where $\hat{J}$ is the rotational kinetic operator and $B$ the
rotational constant.  The excellent agreement between this sudden
impact approximation and the exact time-dependent calculation, with a
root-mean-squared difference less than 0.03, eloquently advocates for
the kick mechanism.

\section{Optimal control scenarios} \label{sec:optimal}

\subsection{General frame}

A completely different approach in its philosophy, without a priori
reference to any dynamical mechanism, is optimal control.  Such a
scheme presents a general structure resting on three steps:
\begin{enumerate}
\item an evaluation function, $\left\langle \cos \theta \right \rangle
  (t)$, involving the numerical solution of the time-dependent
  Schr\"{o}dinger equation (using split
  operator~\cite{split:adb1991,fft:feit82,split:fleck76} or basis
  expansion~\cite{orient:benhajyedder02} techniques).  For a
  laser-driven rigid rotor this amounts to solving
\begin{equation}
i \hbar \frac{\partial}{\partial t} \psi(\theta, \varphi, t) =
\hat{H}(t) \psi(\theta, \varphi, t)
\end{equation}
with 
\begin{equation}
\hat{H}(t) = B \hat{J}^{2} - \mu_{0} \mathcal{E}(t) \cos\theta -
\left[ \left( \alpha_{\parallel} - \alpha_{\perp} \right) \cos^{2}\theta +
    \alpha_{\perp} \right] \frac{\mathcal{E}^{2}(t)}{2}
\end{equation}
in order to calculate
\begin{equation}
\left\langle \cos \theta \right \rangle (t) = \int_{0}^{2\pi}
\int_{0}^{\pi} \left| \psi (\theta, \varphi; t) \right|^{2} \cos\theta
\sin\theta \, d\theta \, d\varphi 
\label{eq:costheta}
\end{equation}

\item a set of parameters defining the temporal shape of the laser
  pulse, given here as a superposition of individual sine-square
  functions with intermediate plateau values,
\begin{equation}
\mathcal{E} (t) = \sum_{n=1}^{N} \mathcal{E}_{n} (t) \sin \left(
  \omega_{n} t + \phi_{n} \right)
\end{equation}
with
\begin{equation}
\mathcal{E}_{n} (t) = \left\{ 
   \begin{array}{llrcl}
      0 & \text{if } & & t & \leq t_{n0} \\
      \mathcal{E}_{n0} \sin^{2} \left[ \frac{\pi}{2} \left( \frac{t -
      t_{n0}}{t_{n1} - t_{n0}} \right) \right] & \text{if } & t_{n0}
      \leq & t & \leq t_{n1} \\  
      \mathcal{E}_{n0} & \text{if } & t_{n1} \leq & t & \leq t_{n2} \\
      \mathcal{E}_{n0} \sin^{2} \left[ \frac{\pi}{2} \left( \frac{t_{n3}
      - t }{t_{n3} - t_{n2}} \right) \right] & \text{if } & t_{n2}
      \leq & t & \leq t_{n3} \\
      0 & \text{if } & & t & \geq t_{n3} 
   \end{array}
\right.
\label{eq:field}
\end{equation}

\item a target $j$, i.e., an optimization criterion.

\end{enumerate}

The optimization aims at the obtainment of the laser parameters
(intensity, frequency, temporal shape defined by rise and fall times,
together with possible plateau durations) that optimally satisfy the
target $j$.  The optimization procedure we use is based on genetic
algorithms~\cite{control:auger02,gp:schoenauer03} operating in a $7N$
dimensional parameter space (i.e., 7 parameters for each of the $N$
individual pulses). It is worthwhile noting that a given criterion
puts the emphasis on a specific feature of the orientation, such that
different criteria may lead to quite different results. This is
basically related to the fact that orientation can by no way be
achieved by monitoring a single quantum state or a unique and
pre-determined superposition of states.  We will analyze in some
detail the consequences of the choice of specific criteria.  But
before doing this, let us proceed to a first optimization retaining an
intuitive, simple criterion that consists in the maximization of
$\left\langle \cos \theta \right\rangle$ for a single time
$t_{\mathrm{f}}$ when the laser field is off,
\begin{equation}
j = \left| \left\langle  \cos \theta \right\rangle (t_{\mathrm{f}})
\right|.
\label{eq:crit0}
\end{equation}

\subsection{The kicked molecule}

\begin{figure}
\includegraphics{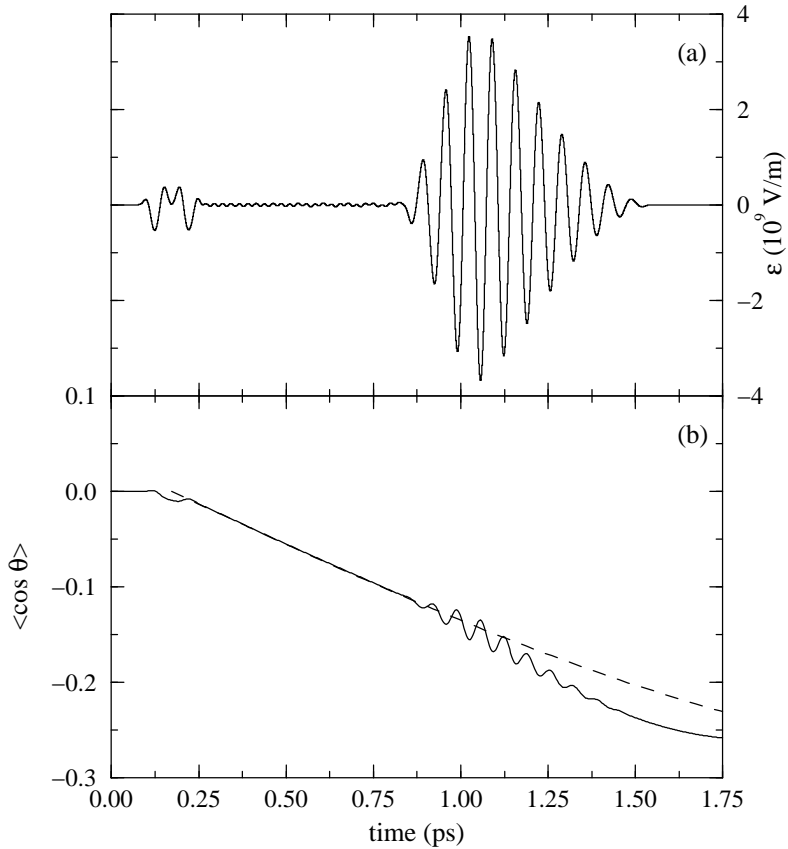}
\caption{(a) Laser field resulting from the optimization using
  criterion $j = \left| \left\langle \cos \theta \right\rangle
    (t_{\mathrm{f}}) \right|$, for the HCN molecule in a rigid-rotor
  model.  (b) Orientation dynamics obtained using this field (solid
  line) and using a sudden-impact approximation (dashed line). (Taken
  from C.~M. Dion, A.~Ben Haj-Yedder, E.~Canc\`{e}s, C.~Le~Bris,
  A.~Keller, and O.~Atabek, Phys. Rev. A \textbf{65} (2002), 063408,
    copyright (2002) by the American Physical Society.)}
\label{fig:crit0}
\end{figure}

\begin{figure}
\includegraphics{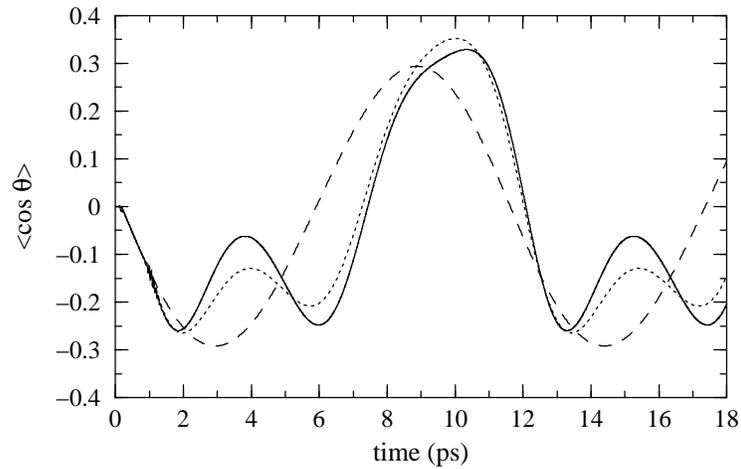}
\caption{Same orientation dynamics as figure~\ref{fig:crit0}(b), but
  given for an entire rotational period of the HCN molecule
  ($T_{\mathrm{rot}} = 11.45\ \mathrm{ps}$) and with an additional
  curve (dotted line) corresponding to the sudden-impact approximation
  with two kicks (see text).  (Taken from C.~M. Dion, A.~Ben
  Haj-Yedder, E.~Canc\`{e}s, C.~Le~Bris, A.~Keller, and O.~Atabek, Phys.
  Rev. A \textbf{65} (2002), 063408, copyright (2002) by the American
  Physical Society.)}
\label{fig:crit0long}
\end{figure}

The optimal electric field which is obtained and illustrated in
figure~\ref{fig:crit0}(a) turns out to be one of the most enlightening
results of this study.
A combination of three individual pulses (i.e., 21 parameters to be
optimized) leads to a very sudden and asymmetric pulse, followed by a
null plateau and some rather symmetrical oscillations.  The molecule
(HCN taken as a rigid rotor, in this example) orients through the
already discussed kick mechanism due to the fast angular momentum
transfer by the unipolar pulse at about $t=0.3\ \mathrm{ps}$ and
during the null plateau lasting 1~ps.  The symmetrical oscillations of
this field between 1 and 1.7~ps do not seem to play a major part in
this dynamics.  Here again, the kick mechanism can be evidenced by a
sudden impact approximation which closely follows the $\left\langle
  \cos \theta \right\rangle$ dynamics at least until 1.5~ps [see
figure~\ref{fig:crit0}(b)].  It is important to note that an accuracy
of 4 digits in the genetic algorithm is achieved for the intensities,
frequencies, and absolute phase differences in order to produce the
particular shape of the electric field of figure~\ref{fig:crit0}(a),
responsible of the subsequent kick mechanism.
Figure~\ref{fig:crit0long} accounts for the long-time dynamics (i.e.,
for full rotational period).
Maximum orientation is reached at about $t
\approx 10\ \mathrm{ps}$ and lasts for about 3~ps.  It is interesting
to note that the result of the sudden impact model based on the first
kick (at $\sim 0.5\ \mathrm{ps}$) can further be improved by
introducing the effect of the oscillatory field acting after the null
plateau as a second kick (that is again freezing the rotational motion
during the time over which the molecule experiences this second
interaction). 

\begin{figure}
\includegraphics{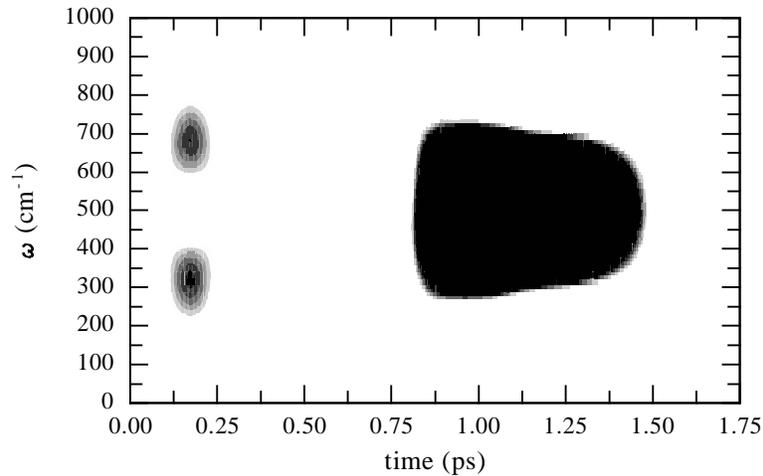}
\caption{Short-time Fourier transform [equation~(\ref{eq:stft})] of
  the field given in figure~\ref{fig:crit0}(a).}
\label{fig:crit0stft}
\end{figure}

An analysis of the optimized laser pulse obtained can be carried out
using the short-time Fourier transform of the electric field, given by
\begin{equation}
\mathcal{F}(\omega,t) = \int_{-\infty}^{+\infty} \mathcal{E}(\tau)
f_{\mathrm{w}}(\tau-t) e^{-i \omega \tau} d\tau,
\label{eq:stft}
\end{equation}
with $f_{\mathrm{w}}$ a windowing function (we have taken a
Tukey-Hanning window~\cite{math:priestley81} of total width 0.29~ps).
The result is shown in figure~\ref{fig:crit0stft}
as a 2D representation of the electric field amplitude as a function
of $\omega$ and $t$.  The asymmetric feature present in the field at
around $\sim 0.15\ \mathrm{ps}$ turns out to be the superposition of
two components at frequencies $\omega$ and $2\omega$.  This
corresponds to the most asymmetrical electric field which, when
truncated at a time that is not an integer of the optical period,
leads to a non-zero field integral imparting the kick to the molecule.
This is followed later on by a component centered around a single
frequency which also leads to a second, smaller kick.  Such a result
has a broader applicability: the same pulse applied to a different
molecule (e.g., LiF) produces similar orientation
dynamics~\cite{orient:dion02}.

\subsection{The optimization criteria}

We now return to a thorough analysis of the consequences of choosing
different criteria in the optimization scheme.  Ideally, the best
orientation is the one which is the most efficient (i.e., with a value
of $\left\langle \cos \theta \right\rangle$ close to one) and which
lasts forever.  The relative merits of the different criteria can thus
be quantitatively measured in terms of the efficiency and the duration
of the orientation.  All calculations that are presented hereafter
apply to the LiF molecule and are conducted within a parameter
sampling space limited to frequencies comprised in between 500 and
$4000\ \mathrm{cm}^{-1}$, to individual pulses extinction times less
than $t_{\mathrm{f}}$ taken as one tenth of a rotational period (as we
are seeking post-pulse orientation resulting from a sudden excitation,
which also helps in keeping calculations within acceptable CPU times)
and to laser intensities not exceeding $3 \times 10^{13}\ 
\mathrm{W/cm}^{2}$ (to avoid risk of ionization damage, having in mind
that $1.1 \times 10^{14}\ \mathrm{W/cm}^{2}$ is the predicted first
ionization threshold for LiF).

\begin{figure}
\includegraphics{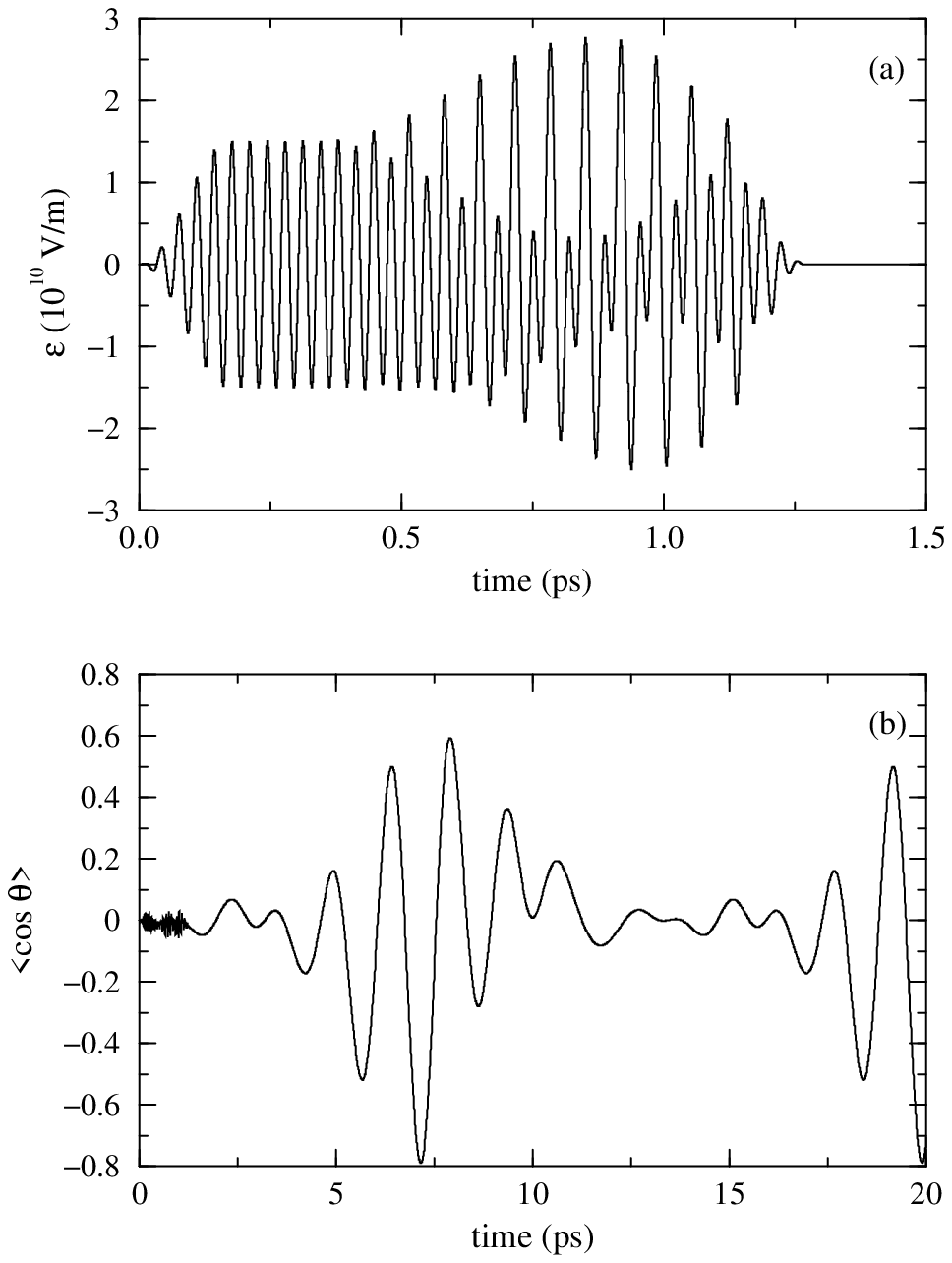}
\caption{(a) Laser field resulting from the optimization using
  criterion $j_{1} = \max_{t \in [t_{f},t_{f}+T_{\mathrm{rot}}]}
  \left| \left\langle \cos \theta \right\rangle (t) \right|$, for the
  LiF molecule in a rigid-rotor model.  (b) Corresponding orientation
  dynamics. (Adapted from A.~Ben Haj-Yedder, A.~Auger, C.~M. Dion,
  E.~Canc\`{e}s, A.~Keller, C.~Le~Bris, and O.~Atabek, Phys. Rev. A
  \textbf{66} (2002), 063401, copyright (2002) by the Americal
  Physical Society.)}
\label{fig:crit1}
\end{figure}

\begin{figure}
  \includegraphics{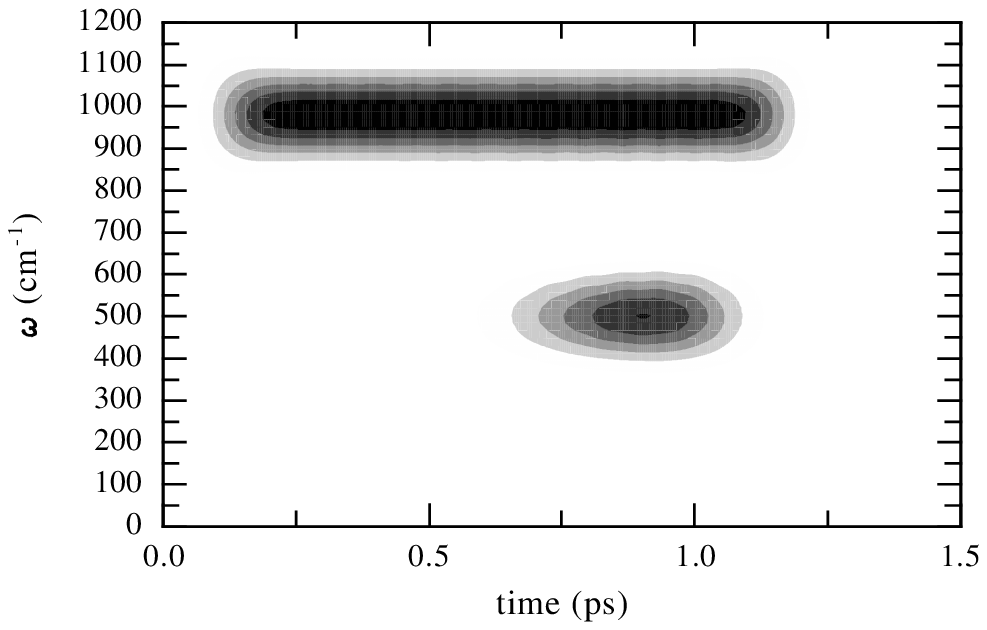}
\caption{Short-time Fourier transform [equation~(\ref{eq:stft})] of the
  field given in figure~\ref{fig:crit1}(a).}
\label{fig:crit1stft}
\end{figure}

The criteria can roughly be classified into two groups: simple, when
they emphasize either efficiency or duration; hybrid, when they
realize a compromise between efficiency and duration.  Aiming at the
best orientation efficiency, a more flexible criterion then the one
that has been used so far [equation~(\ref{eq:crit0})] is
\begin{equation}
j_{1} \equiv \max_{t \in [t_{f},t_{f}+T_{\mathrm{rot}}]}
\left| \left\langle \cos \theta \right\rangle (t) \right|,
\label{eq:j1}
\end{equation}
the maximum being now taken at any time within the interval
$[t_{f},t_{f}+T_{\mathrm{rot}}]$ (taking into account the periodical
behavior of the orientation dynamics with respect to the rotational
period $T_{\mathrm{rot}}$). The result is
displayed in figure~\ref{fig:crit1}.
The electric field thus obtained is built up from two individual
pulses of comparable intensities, with frequencies in a ratio of 2
responsible for the double wiggles in the amplitude, as illustrated in
figure~\ref{fig:crit1stft}.
Excellent orientation ($\left\langle \cos \theta \right\rangle \approx
-0.8$) is achieved with a time delay of about 6~ps after the pulse is
off.  This molecular response time to the laser interaction is the
time needed for the specific phase interferences to occur in the
rotational wave packet.  Such occurrences of the field-free dynamics
are periodical (molecular rotational periodicity $T_{\mathrm{rot}}$),
give rise to the revival structures discussed in the
literature~\cite{align:seideman99}, and are common to all results presented
hereafter.

\begin{figure}
\includegraphics{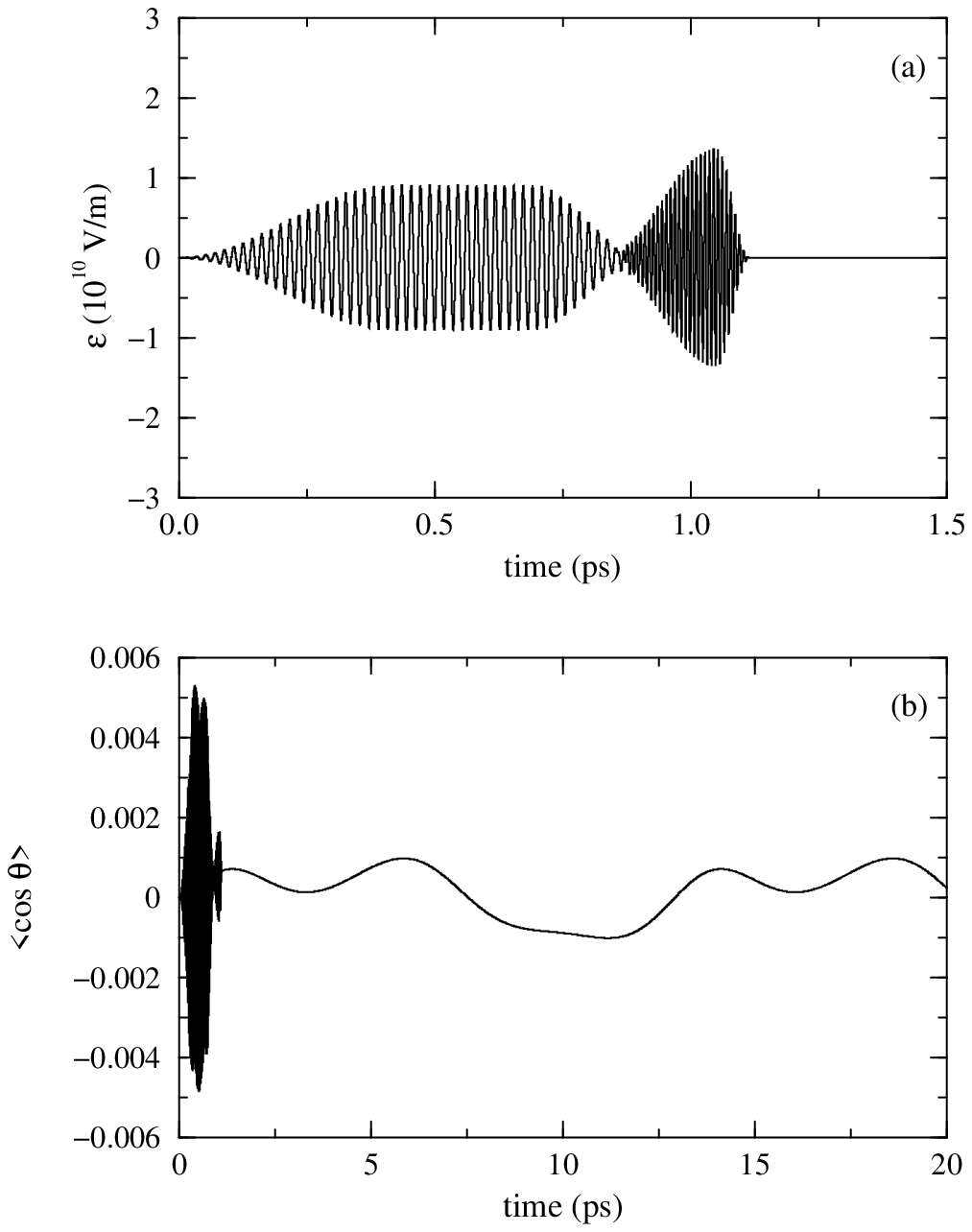}
\caption{(a) Laser field resulting from the optimization using
  criterion $j_{2} = \max_{t \in [t_{f},t_{f}+T_{\mathrm{rot}}]}
  \tau / T_{\mathrm{rot}}$, for the LiF molecule in a
  rigid-rotor model.  (b) Corresponding orientation dynamics. (Adapted
  from A.~Ben Haj-Yedder, A.~Auger, C.~M. Dion, E.~Canc\`{e}s, A.~Keller,
  C.~Le~Bris, and O.~Atabek, Phys. Rev. A \textbf{66} (2002), 063401,
  copyright (2002) by the Americal Physical Society.)}
\label{fig:crit2}
\end{figure}

The criterion $j_{1}$ fails in producing long-lasting orientation (the
time interval over which $\left| \left\langle
\cos \theta \right\rangle \right|$ remains larger than 70\% of its
maximum value does not exceed 0.37~ps).  Another criterion is built to
precisely maximize the orientation duration $\tau$,
\begin{equation}
j_{2} \equiv \max_{t \in [t_{f},t_{f}+T_{\mathrm{rot}}]}
\frac{\tau}{T_{\mathrm{rot}}},
\label{eq:j2}
\end{equation}
where $\tau$ is defined as the time over which an orientation
exceeding $j_{1}/\sqrt{2}$ is kept,
\begin{equation}
  \frac{j_{1}}{\sqrt{2}} \leq \left| \left\langle \cos \theta
  \right\rangle (t) \right| \leq j_{1}.
\end{equation}
The optimized electric field shows two intense, sudden, high frequency
and well separated components (see figure~\ref{fig:crit2}), inducing a
smooth behavior for $\left\langle \cos \theta \right\rangle (t)$ as
expected for large values of $\tau$, but unfortunately the orientation
is very inefficient.

\begin{figure}
\includegraphics{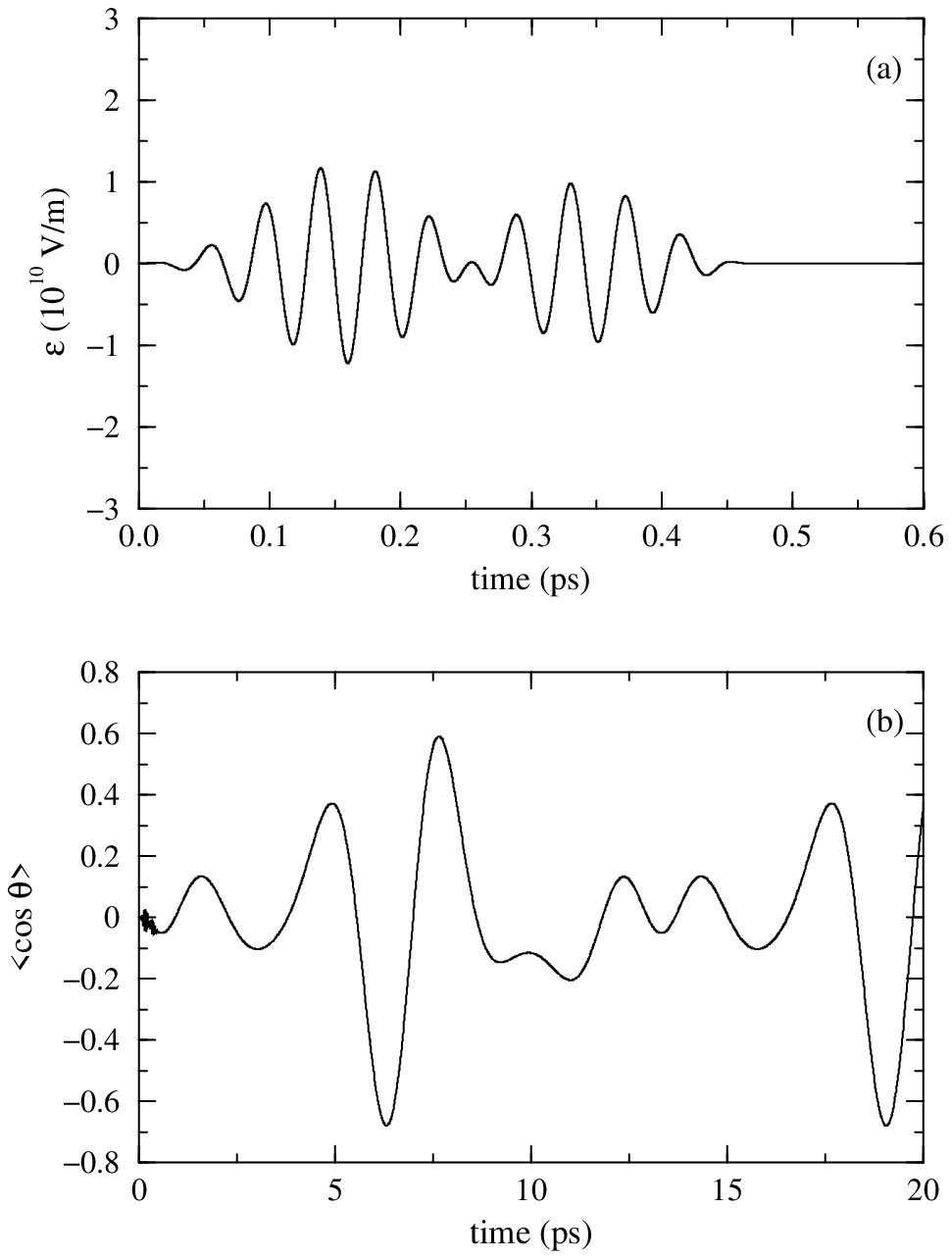}
\caption{(a) Laser field resulting from the optimization using
  criterion $j_{3} = \max \left( 1 / T_{\mathrm{rot}} \right)
  \int_{t_{f}}^{t_{f}+T_{\mathrm{rot}}} \mathcal{C}^{2} (t) \, dt$,
  for the LiF molecule in a rigid-rotor model.  (b) Corresponding
  orientation dynamics. (Adapted from A.~Ben Haj-Yedder, A.~Auger,
  C.~M. Dion, E.~Canc\`{e}s, A.~Keller, C.~Le~Bris, and O.~Atabek, Phys.
  Rev. A \textbf{66} (2002), 063401, copyright (2002) by the Americal
  Physical Society.)}
\label{fig:crit3}
\end{figure}

\begin{figure}
\includegraphics{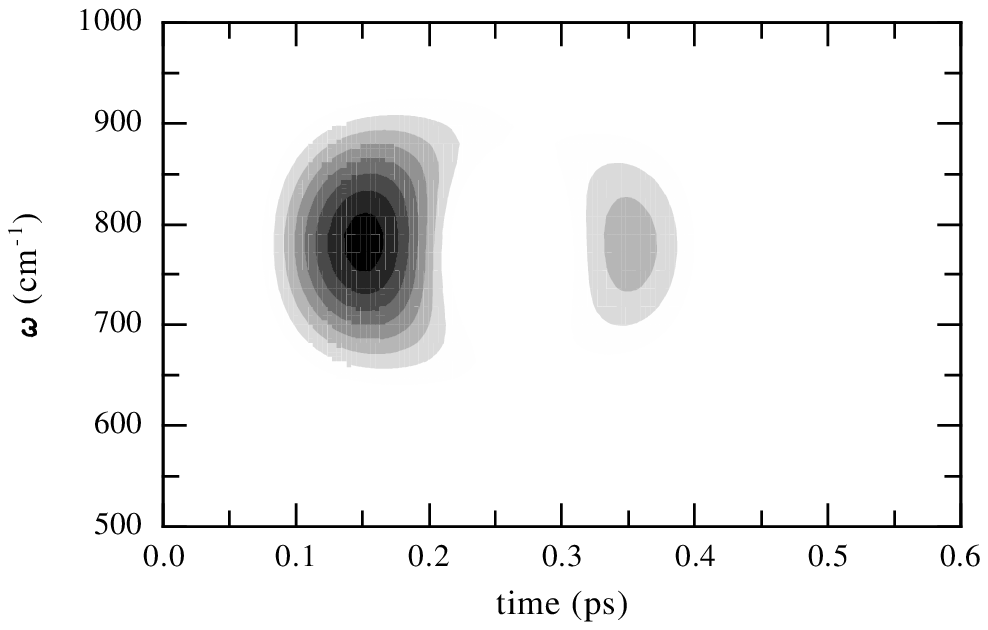}
\caption{Short-time Fourier transform [equation~(\ref{eq:stft})] of
  the field given in figure~\ref{fig:crit3}(a).}
\label{fig:crit3stft}
\end{figure}

The weakness of these simple criteria may be overcome by building
physically more sound hybrid criteria that combine the advantages of
the previous requirements for a better compromise between efficiency
and duration.  Our best result is obtained with a criterion that
maximizes a functional of $\left\langle \cos \theta \right\rangle^{2}
(t)$ over an entire rotational period,
\begin{equation}
  j_{3} \equiv \max \frac{1}{T_{\mathrm{rot}}}
  \int_{t_{f}}^{t_{f}+T_{\mathrm{rot}}} \mathcal{C}^{2} (t) \, dt,
\label{eq:j3}
\end{equation}
where $\mathcal{C}(t)$ is tailored as to put the emphasis on time
intervals where $\left\langle \cos \theta \right\rangle$ is larger
than some fixed value (0.4 in our case) through an appropriate
weighting factor (0.1 in our case):
\begin{equation}
\mathcal{C}(t) = \left\{
\begin{array}{cl}
0.1 \left\langle \cos \theta \right\rangle(t) & \mbox{if } \left\langle
  \cos \theta \right\rangle(t) < 0.4 \\
 \left\langle \cos \theta \right\rangle(t) & \mbox{elsewhere}
\end{array}
\right.
\label{eq:fancycos}
\end{equation}
Figure~\ref{fig:crit3} displays the optimized field built up from two
sudden, intense pulses, with very close frequencies (742~cm$^{-1}$ and
808~cm$^{-1}$) giving a beat-like structure.  It is worthwhile noting
that the resulting field appears as two beats of similar frequency
$\sim 780\ \mathrm{cm}^{-1}$ (see figure~\ref{fig:crit3stft}), but
we failed in analyzing them in terms of a double kick mechanism
(the sudden impact model not providing the expected orientation
dynamics).  Finally, in terms of quantitative measurements, this field
achieves orientation which lasts for 0.7~ps with a $\left|
  \left\langle \cos \theta \right\rangle \right|$ larger than 0.5,
and is one of the best results of the literature.  

\subsection{Robustness with respect to temperature}

\begin{figure}
\includegraphics{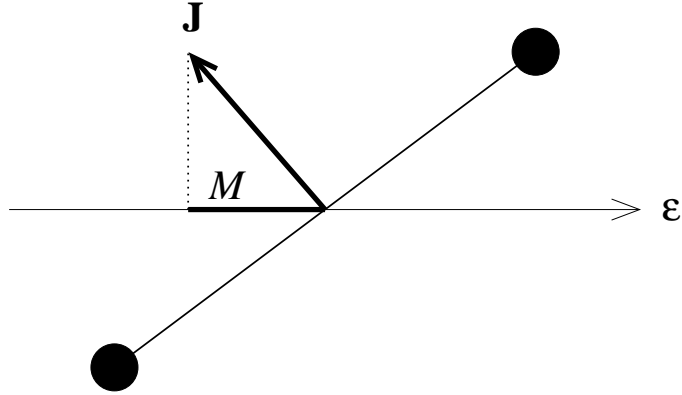}
\caption{Schematic representation of the total angular momentum vector
  $\mathbf{J}$ and its projection $M$ on the laser polarization axis
  $\mathcal{E}$, for a (diatomic) linear molecule. By definition,
  $\mathbf{J}$ is perpendicular to the internuclear axis, therefore if
  $\left| M \right| = J$ the molecule is perpendicular to the laser
  polarization axis.}
\label{fig:jm}
\end{figure}

All calculations so far presented concern an initially rotationless
molecule, $J=M=0$, $M$ being the quantum number labelling the
projection of the rotational angular momentum on the laser
polarization vector taken as the laboratory reference axis.  With
respect to experimental conditions, such a situation concerns a
molecule at a temperature of $T=0\ \mathrm{K}$.  Starting from such a
state (prepared by laser cooling methods, for
instance~\cite{mf:bahns00}), linearly polarized photon absorption only
populate higher $J$ levels; $M$, being a good quantum number
(labelling an azimuthal angular motion, well separated due to
cylindrical symmetry), remains unchanged (i.e., zero). Increasing
temperature, on the other hand, populates higher $J$ levels with all
their $M$ components, $\left| M \right| = 0,1,\ldots,J$, within a
Boltzmann distribution.  For a given component, a vectorial
representation of the field-driven classical rigid rotor leads to
\begin{equation}
\sin \theta = \frac{\left| M \right|}{J},
\end{equation}
the total angular momentum vector $\mathbf{J}$ being orthogonal to the
rigid rotor axis $\mathbf{R}$ and $\theta$ being the polar angle
positioning $\mathbf{R}$ with respect to the laser polarization
vector.  Large values of $M$ would thus prevent alignment and
orientation, $\theta$ reaching a value as high as $\pi/2$ for $\left|
  M \right| = J$, as illustrated
in figure~\ref{fig:jm}.
Using linearly polarized laser sources, no reduction of $M$ could be
achieved, showing that orientation is no more a mathematically
controllable process at non-zero temperature.  Because they are
breaking the overall cylindrical symmetry, elliptically polarized
lasers are possible tools for modifying the values of $M$, in an
optimally controlled way to lead again to alignment and orientation.
Work in this direction is now undertaken in our group.

It remains however that the robustness of our optimal control scheme
can be checked at least for some low temperatures.  Instead of our
previous evaluation function~(\ref{eq:costheta}), we have to refer to
a temperature dependent Boltzmann averaging of $\cos\theta$,
\begin{equation}
\left\langle \left\langle \cos \theta \right\rangle \right\rangle(t)
= Q^{-1} \sum_{J}^{J_{\mathrm{max}}} \exp \left[ \frac{-B
J(J+1)}{k_{B}T} \right] \sum_{M 
  = -J}^{J} \left\langle \cos \theta \right\rangle_{J,M} (t),
\label{eq:cosavgavg}
\end{equation}
where $\left\langle \cos \theta \right\rangle_{J,M} (t)$ results from
the dynamics of an individual initial state $J,M$ included in a
summation over $J$ and $M$ with its appropriate exponential weighting
factor, and a partition function
\begin{equation}
Q = \sum_{J}^{J_{\mathrm{max}}} \left( 2J+1 \right) \exp \left[
\frac{-B J(J+1)}{k_{B}T} 
\right],
\end{equation}
for normalization.

\begin{figure}
\includegraphics{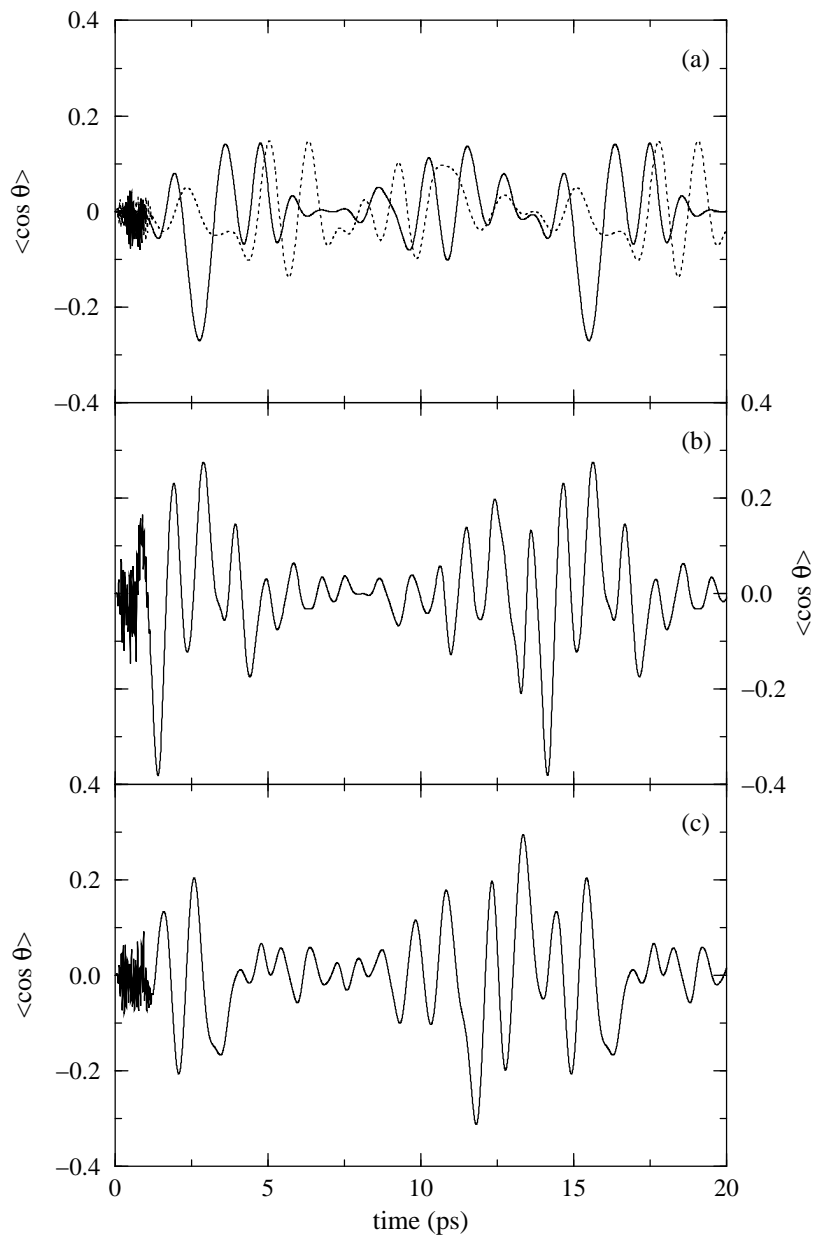}
\caption{Effect of temperature on the orientation dynamics of the LiF
  molecule.  (a) Dotted line: using the laser field obtained from a $T
  = 0\ \mathrm{K}$ optimization with criterion $j_1$ [see
  figure~\ref{fig:crit1}(a)], but applied to an initial distribution
  of molecules at $T = 5\ \mathrm{K}$.  Solid line: using the laser
  field obtained from a $T = 5\ \mathrm{K}$ optimization with
  criterion $j_1$.  (b) Same as (a), but using a superposition of 3
  laser pulses.  (c) Same as (b), but using criterion $j_3$. (Adapted
  from A.~Ben Haj-Yedder, A.~Auger, C.~M. Dion, E.~Canc\`{e}s, A.~Keller,
  C.~Le~Bris, and O.~Atabek, Phys. Rev. A \textbf{66} (2002), 063401,
  copyright (2002) by the Americal Physical Society.)}
\label{fig:temp}
\end{figure}

Figure~\ref{fig:temp} presents the results for the LiF molecule at
$T=5\ \mathrm{K}$.
Here again, simple ($j_1$) and hybrid ($j_3$-like) criteria are
referred to.  The dotted line (panel a) represents the orientation
dynamics at 5~K [equation~(\ref{eq:cosavgavg})] but with the electric
field of figure~\ref{fig:crit1}(a) optimized for $T=0\ \mathrm{K}$.
This is, precisely, to show how orientation effects are rapidly washed
out when incorporating higher initial $M$'s.  A field optimized for
$T=5\ \mathrm{K}$ restitutes again the oscillatory structures of
$\left\langle \left\langle \cos \theta \right\rangle \right\rangle
(t)$, with however reduced amplitudes (solid line of panel a).  A more
satisfactory result requires the introduction of an additional
individual pulse bringing 21 free parameters in the optimization
scheme.  The result in given in figure~\ref{fig:crit1}(b) with an
orientation higher than $\left\langle \left\langle \cos \theta
  \right\rangle \right\rangle = 0.27$ kept over a time of about
$0.25$~ps.  The result of the hybrid criterion $j_3$ with a
21-free-parameters optimization is illustrated in
figure~\ref{fig:crit1}(c).  A much longer time duration ($\sim 0.4\ 
\mathrm{ps}$) is achieved for about the same efficiency.

\section{Conclusion} \label{sec:conc}

The complementarity between basic mechanisms and optimal control
schemes can be summarized either by referring to the way of
implementing  mechanisms in control, or reversely of identifying
mechanisms revealed from control.

The first strategy rests on three observations:
\begin{enumerate}
\item there is no single solution arising from optimal control
  scenarios (not only different criteria, but also different sampling
  spaces for parameters would lead to different results);

\item a careful study of the laser-induced dynamics can help to the
  identification (or guess) of some basic mechanisms leading to the
  desired observable;

\item by appropriately building the targets and delineating the
  parameter sampling space, we can help the optimal control scheme to
  take advantage of the mechanisms.
\end{enumerate}

The second strategy consists in learning a dynamical mechanism
presenting some robustness from the optimal control results.  This
mechanism may be used directly for reaching the desired observable
(strategy 2) or further improved by implementing it again in a control
scenario with some additional flexibility (strategy 1).

The kick mechanism and the optimal control carried out using sudden
pulses is an example that illustrates these two routes.  The mechanism
has actually been depicted using half-cycle pulses and a sudden impact
model.  The parameter sampling space for the optimal control, although
not constrained, has been limited to sudden pulses that are precisely
compatible with the kick mechanism.  This control scenario has finally
revealed a double-kick mechanism which can serve as a starting point
of a train of kicks, that we can merely guess to provide even better
orientation.  Finally, this mechanism, together with some additional
free parameters such as the time delay between the successive kicks,
can be re-injected (re-introduced) in an optimal control scheme
(strategy 1) aiming at improving its efficiency.  We are presently
pursuing some investigations in this direction.


\providecommand{\bysame}{\leavevmode\hbox to3em{\hrulefill}\thinspace}
\providecommand{\MR}{\relax\ifhmode\unskip\space\fi MR }
\providecommand{\MRhref}[2]{%
  \href{http://www.ams.org/mathscinet-getitem?mr=#1}{#2}
}
\providecommand{\href}[2]{#2}

\end{document}